\documentclass[preprint]{aastex62}

\newcommand{\speed}[1]{#1 km~s${}^{-1}$}
\newcommand{\accel}[1]{#1 km~s${}^{-2}$}
\newcommand{\acc}[1]{#1 m~s${}^{-2}$}
\newcommand{\nfig}[1]{Figure~\ref{#1}}

\usepackage{amsmath}

\received{... ..., 2018}
\revised{... ..., 2018}
\accepted{... ..., 2018}
\submitjournal{ApJ}

\shorttitle{First unambiguous large-scale quasi-periodic EUV wave}
\shortauthors{Shen et al.}

\begin{document}

\title{First Unambiguous Imaging of Large-Scale Quasi-Periodic Extreme-Ultraviolet Wave or Shock}

\correspondingauthor{Yuandeng Shen}
\email{ydshen@ynao.ac.cn}

\author[0000-0001-9493-4418]{Yuandeng Shen}
\affiliation{Yunnan Observatories, Chinese Academy of Sciences,  Kunming, 650216, China}
\affiliation{State Key Laboratory of Space Weather, Chinese Academy of Sciences, Beijing 100190, China}
\affiliation{Center for Astronomical Mega-Science, Chinese Academy of Sciences, Beijing, 100012, China}
\author[0000-0002-7289-642X]{P. F. Chen}
\affiliation{School of Astronomy \& Space Science, Nanjing University, Nanjing 210023, China}
\affiliation{Key Laboratory of Modern Astronomy and Astrophysics, Nanjing University, Nanjing 210023, China}
\author[0000-0002-3483-5909]{Ying D. Liu}
\affiliation{State Key Laboratory of Space Weather, Chinese Academy of Sciences, Beijing 100190, China}
\affiliation{University of Chinese Academy of Sciences, Beijing, China}
\author{Kazunari Shibata}
\affiliation{Kwasan and Hida Observatories, Kyoto University, Yamashina-ku, Kyoto 607-8471, Japan}
\author{Zehao Tang}
\affiliation{Yunnan Observatories, Chinese Academy of Sciences,  Kunming, 650216, China}
\affiliation{University of Chinese Academy of Sciences, Beijing, China}
\author[0000-0002-7694-2454]{Yu Liu}
\affiliation{Yunnan Observatories, Chinese Academy of Sciences,  Kunming, 650216, China}
\affiliation{Center for Astronomical Mega-Science, Chinese Academy of Sciences, Beijing, 100012, China}

\begin{abstract}
We report the first unambiguous quasi-periodic large-scale extreme-ultraviolet (EUV) wave or shock that was detected by the Atmospheric Imaging Assembly on board the {\em Solar Dynamics Observatory}. During the whip-like unwinding eruption of a small filament on 2012 April 24, multiple consecutive large-scale wavefronts emanating from AR11467 were observed simultaneously along the solar surface and a closed transequatorial loop system. In the meantime, an upward propagating dome-shaped wavefront was also observed, whose initial speed and deceleration are about \speed{1392} and \accel{1.78}, respectively. Along the solar surface, the quasi-peridoic wavefronts had a period of about 163$\pm$21 seconds and propagated at a nearly constant speed of \speed{747$\pm$26}; they interacted with active region AR11469 and launched a sympathetic upward propagating secondary EUV wave. The wavefronts along the loop system propagated at a speed of \speed{897}, and they were reflected back at the southern end of the loop system at a similar speed. In addition to the propagating waves, a standing kink wave was also present in the loop system simultaneously. Periodicity analysis reveals that the period of the wavefronts was consistent with that of the unwinding helical structures of the erupting filament. Based on these observational facts, we propose that the observed quasi-periodic EUV wavefronts were most likely excited by the periodic unwinding motion of the filament helical structures. In addition, two different seismological methods are applied to derive the  magnetic field strength of the loop system, and for the first time the reliability of these inversion techniques are tested with the same magnetic structure.
\end{abstract}


\section{Introduction}
Large-scale wave phenomena in the solar atmosphere were firstly detected in the dense chromosphere, being observed as propagating arc-shaped dark/white fronts in the H$\alpha$ off-band filtergrams. They were later called Moreton waves in history \citep[e.g.,][]{1960AJ.....65U.494M,2007ApJ...658.1372B,2008ApJ...685..629G,2008ApJ...684L..45N,2010ApJ...708.1639M,2013ApJ...773..166L}. Moreton waves can propagate to a distance of the order of 1000 Mm with a speed of the order \speed{1000} \citep[e.g.,][]{1961ApJ...133..935A,1964AJ.....69Q.145M,2012ApJ...752L..23S}, and they are often associated with flares, coronal mass ejections (CMEs), and type II radio bursts. Since the Alfv\'en speed in the chromosphere is of the order of \speed{100}, a shock wave with an  Alfv\'en Mach number of about 10, if Moreton waves are really propagating waves in the chromosphere, would be completely dissipated before having propagated over 1000 Mm. Therefore, it is hard to explain the existence of Moreton waves in the dense chromosphere. To interpret Moreton waves, \cite{1968SoPh....4...30U} proposed a theoretical model in which Moreton waves are the imprints as coronal fast-mode waves or shocks sweep the chromosphere. This scenario implies that there must be counterparts of Moreton waves in the upper corona, and they should have similar speeds and morphologies. Similar large-scale wavelike disturbances were discovered by the EUV Imaging Telescope (EIT) aboard {\em SOHO} \citep{1997SoPh..175..571M,1998GeoRL..25.2465T}, which were called ``EIT waves" initially, and more often called extreme-ultraviolet (EUV) waves later. In addition, the coronal counterparts of Moreton waves were also confirmed by soft X-ray and \ion{He}{1} 10830 \AA\ observations \citep[e.g.,][]{2002AA...383.1018K,2002A&A...394..299V,2002ApJ...572L.109N,2004PASJ...56L...5N,2004ApJ...610..572G,2005ApJ...626L.121W}.

The driving mechanism and physical nature of large-scale EUV waves were hotly debated in the past twenty years \citep{2017SCPMA..60b9631C}. Because chromospheric Moreton waves were believed to be driven by the flare pressure pulses for a long time since \citet{1968SoPh....4...30U}, solar physicists naturally thought at the beginning that EUV waves also have the same driving mechanism \citep[e.g.,][]{2002AA...383.1018K,2003SoPh..212..121H,2004AA...418.1117W}. However, more and more recent high-resolution observational studies have shown that EUV waves are in fact bow shocks driven by CMEs instead of flare pressure pulses \citep[e.g.,][]{2009ApJ...700L.182P,2009ApJ...703L.118K,2010A&A...522A.100P,2006ApJ...641L.153C,2012ApJ...754....7S,2012ApJ...752L..23S,2017ApJ...851..101S,2013AA...556A.152X,2017SoPh..292....7L,2018A&A...612A.100C,2011ApJ...734...84L}. Further investigations indicated that EUV waves can also be excited by other kinds of slightly different processes, such as newly-formed expanding loops, coronal jets, and mini-filament eruptions in small scales \citep[e.g.,][]{2010ApJ...709..369P,2012ApJ...753L..29Z,2013MNRAS.431.1359Z,2015ApJ...804...88S,2017ApJ...851..101S}, and the sudden expansion of transequatorial loops driven by coronal jets in large scales \citep{2018ApJ...860L...8S,2018MNRAS.480L..63S} . In these scenarios, those authors found that such kind of EUV waves often have a shorter lifetime (a few minutes) compared to those driven by CMEs (tens of minutes). Moreover, \cite{2018ApJ...861..105S} reported an interesting non-CME-associated homologous EUV wave event in which the initial speeds of the waves are about \speed{1000}, and the authors proposed that the observed EUV waves were large-amplitude nonlinear fast-mode magnetosonic waves or shocks directly driven by the associated recurrent coronal jets, resembling the generation of a piston shock in a tube. These new observations suggest that EUV waves can be excited by different kinds of coronal disturbances, e.g., CMEs, coronal jets, and mini-filament eruptions.

Owing to the similar morphology and nearly cospatial relationships between some coronal EUV waves and chromospheric Moreton waves, the former is considered as the coronal counterpart of the latter \citep[e.g.,][]{2000SoPh..193..161T,2001ApJ...560L.105W,2006ApJ...647.1466V,2008SoPh..253..263G,2010ApJ...708.1639M,2012ApJ...745L..18A,2012ApJ...752L..23S}. Regarding the physical nature of EUV waves, they were firstly interpreted as fast-mode magnetosonic waves \citep[e.g.,][]{2000ApJ...543L..89W,2001JGR...10625089W,2002ApJ...574..440O}. However, some characteristics of EUV waves can not be explained in terms of MHD waves, such as their low speeds (i.e., below  \speed{150} in the quiet corona) and the appearance of stationary brightenings. This led to the proposals of various non-wave models of EUV waves, such as the magnetic field-line stretching model \citep[e.g.,][]{2002ApJ...572L..99C,2005ApJ...622.1202C}, the current shell model \citep[e.g.,][]{2000ApJ...545..512D,2008SoPh..247..123D}, and the reconnection front model \citep{2007ApJ...656L.101A}. In particular, \citet{2002ApJ...572L..99C,2005ApJ...622.1202C} proposed that there should be two types of EUV waves, a faster one that corresponds to the coronal counterpart of Moreton wave, and a slower one that is generated by the successive stretching of the closed field lines overlying the erupting flux rope. According to this model, the slower EUV wave has a speed of about three times slower than that of the faster one if the coronal magnetic loops are something like semicircles, and this scenario has been confirmed by many observations \citep[e.g.,][]{2004A&A...427..705Z,2011ApJ...732L..20C,2012ApJ...752L..23S,2013ApJ...773L..33S,2014ApJ...786..151S,2012ApJ...753...52L,2012ApJ...745L..18A,2012ApJ...745L...5C,2013AA...553A.109K,2018ApJ...863..101C}. It seems that the diverse features of EUV waves can only be explained by such a hybrid model in which the fast and slow components are respectively interpreted as fast-mode magnetosonic wave and non-wave features \citep[e.g.,][]{2002ApJ...572L..99C,2009ApJ...705..587C,2012ApJ...750..134D}. It should be noted that the slow component of EUV waves is not a real MHD wave in several models mentioned above, however, some authors also proposed that the slow wave component can be regraded as slow-mode magnetosonic waves \citep[e.g.,][]{2007ApJ...664..556W, 2009ApJ...700.1716W,2012SCPMA..55.1316M}. In addition, the slow component could also be a signature of the laterally expanding CME flanks, i.e., the plasma pileup in front of the expanding flux rope \citep[e.g.,][]{2004A&A...427..705Z,2010A&A...522A.100P}. More details about EUV wave can be found in recent review papers \citep[e.g.,][and references therein]{2014SoPh..289.3233L,2015LRSP...12....3W, Chen16}.

Despite the debate on the nature of slow component of EUV waves, it is widely accepted that the fast component of EUV waves, especially when faster than \speed{500}, is a fast-mode magnetoacoustic wave or shock wave \citep{2012ApJ...745L...5C,2017SoPh..292....7L}. Their wave nature can be inferred by various observational features, such as reflection \citep[e.g.,][]{2008ApJ...680L..81L, 2009ApJ...691L.123G, 2012ApJ...752L..23S, 2012ApJ...746...13L, 2012ApJ...756..143O, 2013ApJ...775...39Y}, transmission \citep[e.g.,][]{2012ApJ...753...52L,2012ApJ...756..143O,2013ApJ...773L..33S}, refraction \citep[e.g.,][]{2012ApJ...754....7S,2013ApJ...773L..33S}, and mode conversion \citep[e.g.,][]{2016SoPh..291.3195C, 2017ApJ...834L..15Z,2018ApJ...863..101C}. In particular, \cite{2013ApJ...773L..33S} observed many of these features in a single EUV wave event, which clearly demonstrated the wave nature of EUV waves.  Another feature associated with the fast-component EUV waves is the appearance of multiple wavefronts. However, so far the direct imaging of such kind of evidence is still very scarce besides some possible signals of multiple wavefronts such as ripple-like features  \citep{2003SoPh..212..121H,2010ApJ...723L..53L}.

In this paper, we present an intriguing EUV wave event on 2012 April 24, which was accompanied by a {\em GOES} C3.7 flare, the eruption of an unwinding filament, a partial halo CME, a chromospheric Moreton wave, and a type II radio burst. It is observed for the first time that several phenomena were all involved in a single EUV wave event, such as quasi-periodic multiple fast-component EUV wavefronts, channeled propagation along a closed coronal loop and its reflection, and so on. The {\it SDO}/AIA data with unprecedented resolution provide an opportunity to understand the driving mechanism of the multiple quasi-periodic EUV wavefronts and the other phenomena that launched by the interaction of the EUV wave with remote magnetic structures. Instruments and observations are briefly introduced in Section \ref{sect2}; Main analysis results are presented in Section \ref{sect3}; Coronal seismology applications are given in Section \ref{sect4}; The last section includes the Discussion and Conclusion of the present study.

\section{Instruments and Observations}\label{sect2}
The event was simultaneously observed by the {\em Solar Dynamics Observatory} \citep[{\em SDO};][]{2012SoPh..275....3P} and {\em Solar Terrestrial Relations Observatory} \citep[{\em STEREO},][]{2008SSRv..136....5K} Behind ({\em STEREO}-B) from two different view angles. On 2012 April 24, the separation angle between the two satellites was about $118^{\circ}$. The Atmospheric Imaging Assembly \citep[AIA;][]{2012SoPh..275...17L} on board {\em SDO} provides continuous full-disk images of the chromosphere and the corona in seven EUV and three UV-visible channels; the pixel size of the AIA images is $0\farcs 6$, and the EUV and UV-visible images have 12 and 24 s cadences, respectively. The full-disk 195 \AA\ and 304 \AA\ images taken by {\em STEREO}-B are used as well, and their cadences are 5 and 10 minutes, respectively. The pixel size of the {\em STEREO}-B images is $1\farcs 59$. The H$\alpha$ line center images were taken by the Chromospheric Telescope \citep[ChroTel;][]{2008SPIE.7014E..13K,2011A&A...534A.105B}, whose time cadence and pixel size are about 3 minutes and $1\arcsec$, respectively. The radio dynamic spectra was provided by the Rosse Solar-Terrestrial Observatory \citep[RSTO;][]{2012SoPh..280..591Z}. We use the low receiver that operates at 10--100 MHz and takes four spectra per second and with 200 channels. In addition, the X-ray fluxes recorded by the {\em Geostationary Operational Environmental Satellite} ({\em GOES}) and {\em Reuven Ramaty High Energy Solar Spectroscopic Imager} \citep[{\em RHESSI};][]{2002SoPh..210....3L} are also used to analyze the flare. In this paper, a running difference (ratio) image is created by subtracting (dividing) a direct image by the previous one; a base ratio image is obtained by dividing images by a fixed image before the flare.

\section{Observational Results}\label{sect3}
On 2012 April 24, a {\em GOES} C3.7 flare occurred in AR11467 (N14, E58), whose start, peak, and end times were at about 07:38, 07:45, and 08:00 UT, respectively. The flare was accompanied by the eruption of an unwinding filament, a partial halo CME, and quasi-periodic large-scale arc-shaped wavefronts propagating simultaneously along the solar surface and a closed transequatorial loop system that connects AR11467 and AR11469 in opposite hemispheres. The wavefronts further caused transverse oscillations of the transequatorial loop system and a large remote quiescent filament on the northwest direction of AR11467. Based on the measurement of the online CDAW (Coordinated Data Analysis Workshops) CME catalog\footnote{\url{https://cdaw.gsfc.nasa.gov/CME_list/index.html}},  the average speed and deceleration of the CME were about \speed{443} and \acc{-8.1}, respectively. Here, the speed (deceleration) of the CME was obtained by fitting the height-time data points of the CME leading front with a linear (second degree polynomial) function. The top row of \nfig{fig1} shows the pre-eruption configuration with AIA 171 \AA\, 304 \AA\ and ChroTel H$\alpha$ line center images. The active region filament and the transequatorial loop system can well be identified in \nfig{fig1}(a), and the large quiescent filament was on the north of AR11467 (see \nfig{fig1}(b) and (c)). \nfig{fig1}(d) shows the {\em GOES} and {\em RHESS} X-ray fluxes, while \nfig{fig1}(e) shows the flare lightcurves in the eruption source region at AIA's different wavelength channels. It is noted that before the impulsive phase of the flare there is a small bump (peaked at about 07:38 UT) in the {\em RHESSI} 6--12, and 12--25 keV energy bands and the AIA lightcurves.

The eruption of the filament in AR11467 is shown in \nfig{fig2} with AIA and {\em STEREO}-B 304 \AA\ images from two different view angles (see also the animation available in \nfig{fig2}). The filament eruption was close to the east (west) limb of the solar disk in the FOV of {\em SDO} ({\em STEREO}-B). The southern part of the filament firstly erupted at about 07:00:00 UT, which showed obvious rotation motion and the eruption direction was in the southeast in SDO/AIA and southwest in STEREO-B/EUVI. The 304 \AA\ images at 07:32:32 UT (AIA) and 07:36:15 UT ({\em STEREO}-B) are plotted in \nfig{fig2} (a) and (d) to show the erupting filament, whose eruption direction is indicated by the blue arrows. This eruption caused the small bump observed in the flare lightcurves and X-ray fluxes (see \nfig{fig1}). The main body of the active region filament began to rise at about 07:37:20 UT and erupted violently at about 07:40:56 UT, which was associated with the main peak in the flare curves; During this process, the whip-like eruption of the filament showed strong unwinding and  lateral expanding motions. The start of the violent eruption of the filament main body is shown in \nfig{fig2} (b) and (e). The unwinding structure is shown in \nfig{fig2} (c) and (f), in which one can easily identify the helical structure of the erupting filament. It is identified that the unwinding direction of the filament was clockwise when viewed from above along the filament axis (see the curved arrows in \nfig{fig2} (c) and (f)). The eruption of the filament showed a jet-like shape and with strong unwinding motion. This morphology resembles the eruption process of unwinding coronal jets and suggests the releasing of the magnetic twist stored in the filament into the outer corona \citep{1986SoPh..103..299S,1987SoPh..108..251K,2011ApJ...735L..43S,2012RAA....12..573C,2013RAA....13..253H,2018ApJ...869...39M}. In recent years, more and more high-resolution observations have indicated that amounts of coronal jets are in fact dynamically related to the eruption of mini-filaments confined by the jet-base, and this kind of jets are called as blowout jets \citep[e.g.,][]{2010ApJ...720..757M,2012ApJ...745..164S,2017ApJ...851...67S,2016ApJ...821..100S,2018MNRAS.476.1286J}. Therefore, the present filament eruption could be regarded as a so-called blowout jet in the knowledge framework of coronal jets.
 
The violent eruption of the main filament body launched striking multiple large-scale arc-shaped wavefronts simultaneously along the solar surface and the transequatorial loop system, and an upward propagating dome-shaped wavefront ahead of the erupting filament. The evolution and morphology of the waves are displayed in \nfig{fig3}. The first EUV wavefront appeared at about 07:41:07 UT, which was about 10 seconds after the start time of the violent eruption of the main filament eruption (see the animation available in \nfig{fig3}). The wavefront initially had a dome-shaped structure off the solar limb and a semicircular shape along the solar surface (see \nfig{fig3}(a)). It is interesting that several EUV wavefronts followed the first one successively along the solar surface (see \nfig{fig3}(b) and (c)). The appearance times of the second and third wavefronts were respectively about 07:45:31 UT and 07:47:31 UT in the AIA 193 \AA\ images. Therefore, the average time lag between the different wavefronts was about 3 minutes. It is noted here that only the fast-component EUV waves were detected in this event. The slow-component EUV wave was not detected at least in our chosen direction. When the EUV wave propagated close to AR11469, the eastern part of the wavefront propagated faster than its western part so that the wave front became significantly deformed (see the wavefront highlighted by the blue dashed curves in \nfig{fig3}(b) and (c)). The deformation of the wavefront was probably caused by the stronger magnetic field strength in AR11469 than the ambient quiet-Sun, because stronger magnetic field strength implies a faster Alfv\'{e}n speed and therefore a faster fast-mode magnetosonic wave speed. This phenomenon also implies that these EUV wavefronts are fast-mode waves (or shock waves). Quasi-periodic wavefronts were also observed in the northwest direction of AR11467, but they did not propagate to a large distance like those in the southwest direction. This was probably due to the blocking of the magnetic structure hosting the large quiescent filament which showed obvious transverse oscillation after the arrival of the EUV wave. The observed quasi-periodic wavefronts can be observed in all AIA EUV channels. In addition, the intensity increase of the first three EUV wavefronts are in the range of 10\%--35\%. This value is consistent with those measured in typical EUV waves \citep{2015LRSP...12....3W}, but much higher than those (1\%--5\%) detected in quasi-periodic fast-propagating magnetosonic waves that are mainly visible in the AIA 171 \AA\ and sometimes 193 \AA\ wavelengths and often along open or closed coronal loops \citep[e.g.,][]{2011ApJ...736L..13L,2012ApJ...753...52L,2012ApJ...753...53S,2013SoPh..288..585S,2018ApJ...853....1S,2018MNRAS.477L...6S,2018MNRAS.480L..63S}. Therefore, we interpret the observed multiple EUV wavefronts as quasi-periodic large-scale EUV waves rather than locally propagating quasi-periodic fast-propagating magnetosonic waves. Despite the similar physical nature of the two kinds of EUV waves, their geometric scales, amplitudes, and generation mechanisms are largely different. 

The interaction between the wave and AR11469 launched a small-scale plasma ejecta seen in the AIA 304 \AA\ observations (see \nfig{fig2} (g) and (h)) and a secondary EUV wave that propagated in the southeast direction above AR11469. This secondary wave was very weak so that it is hard to distinguish directly in static images, but it is clear in the animation made from AIA 193 \AA\ running ratio images. Due to the close temporal and spatial relationships between the arrival of the EUV wave and the occurrence of the plasma ejecta and the secondary EUV wavefront, we propose that the later activities were directly launched by the interaction of the EUV wave with AR11469, and this also indicates that EUV waves can be a good physical linkage for launching sympathetic solar activities \citep[see also,][]{2000GeoRL..27.1083K,2014ApJ...786..151S,2014ApJ...795..130S,2019ApJ...870...15L,2018ApJ...864L..24L}.

It is interesting that quasi-periodic wavefronts are also observed simultaneously along the transequatorial loop system (see the white dashed curves in \nfig{fig3} (e)). The wavefronts originated from AR11467 and propagated along the transequatorial loop system. When reaching the southern end of the loop system, they were reflected back (see the arrow in \nfig{fig3}(f)). The reflected wavefronts propagated from AR11469 along the transequatorial loop system back to AR11467, which can be well observed in the online animation associated with this figure. Similar reflection of fast-mode EUV wave was also previously reported in \cite{2015ApJ...803L..23K}. However, they observed only one wavefront, which was interpreted to be generated by an impulsive energy release.  In addition, the transequatorial loop system showed obvious standing kink oscillations during the propagation of the wavefronts along it.

The EUV wave was also accompanied by a chromospheric Moreton wave that can well be identified in the running difference ChroTel H$\alpha$ images (see \nfig{fig3} (g)). The Moreton wave is highlighted by the red dashed curve, and it had a similar shape and was nearly cospatial with the first EUV wavefront, consistent with the so far well-established scenario that fast-component EUV waves are indeed the counterparts of the chromospheric Moreton waves \citep[e.g.,][]{2000SoPh..193..161T,2001ApJ...560L.105W,2012ApJ...752L..23S}. In the {\em STEREO}-B 195 \AA\ running difference images, the dome-shaped wavefront over the limb can be clearly observed (see the white dotted curve in \nfig{fig3} (h)). However, the wavefronts along the transequatorial loop system and the solar surface were hard to distinguish. In addition, a well defined type II radio burst was recorded by the RSTO spectrometer. The start time of the  fundamental band of the type II radio burst was at 07:48 UT, and the start frequency was at about 42 MHz. Based on the relationship between plasma frequency and density \citep{1985srph.book.....M}, and the empirical coronal electron density model \citep{1999ApJ...523..812S} as used in many articles \citep[e.g.,][]{2006ApJ...649.1110L,2011ApJ...738..160M,2016SoPh..291.3369G}, we can derive the height and therefore the velocity of the corresponding shock wave. Here, we use the center frequency data points (indicated by the white dashed line in \nfig{fig3} (i)) as the input in our calculation; it is obtained that the velocity of the shock was about \speed{615} and its start height was about 0.51 solar radius above the solar surface. The shock speed and height derived from the type II burst are similar to those obtained by \cite{2009ApJ...691L.151L} for the 2007 December 31 eruption.

Time-distance stack plots are created for analyzing the kinematics of the quasi-periodic EUV waves, oscillations of the transequatorial loop and the large quiescent filament, and the eruption of the unwinding filament in the source region. To obtain a time-distance stack plot, we first obtained the one-dimensional intensity profiles along a specified path at different times, and then a two-dimensional time-distance stack plot was generated by stacking the obtained one-dimensional intensity profiles in time. \nfig{fig4}(a)--(f) are the time-distance stack plots made along paths L1--L6 as shown in \nfig{fig1}(a)--(c), respectively. 

As shown in \nfig{fig4}(a), the quasi-periodic EUV wavefronts along the solar surface can well be identified as bright inclined stripes; By fitting each of the first three strong stripes with a second quadratic function, it is obtained that the decelerations of the three wavefronts are similar, and their average value is about \accel{0.36}. To obtain the average speed of the wavefronts, we simply fit each of the stripes with a linear function since the wavefronts showed a nearly constant speed. The result indicates that the average speed of first three wavefronts was about \speed{747 $\pm$ 26}. The upward propagating dome-shaped EUV wave is also measured (not shown here). The result indicates that the initial speed was about \speed{1392}, and the deceleration was about \accel{1.78}. This suggests that the upward propagating EUV wave was much faster than the lateral propagating component along the solar surface, consistent with the result presented in \cite{2010ApJ...716L..57V}. As described above, the appearance of the type II radio burst was delayed with respect to the start time of the EUV wave along the solar surface by about seven minutes. Based on the equation $\rm v = v_{0} + at$ and the measured initial speed and deceleration of the upward propagating dome-shaped wave originated from AR11467, it is derived that the speed of the wave should be \speed{644} seven minutes after its first appearance. This result is in agreement with the wave speed derived from the type II radio burst (\speed{615}), and it also suggests that the observed type II radio burst should be caused by the observed EUV wave.

The secondary wave above AR11469 showed obvious deceleration (see \nfig{fig4} (b)). Measurement indicates that the initial speed (within 100--200 Mm) and deceleration were about \speed{1358} and \accel{4.5}, respectively. It should be pointed out that the speed and deceleration of the secondary wave have relatively large errors, because this wave was very weak so that it is hard to trace.

The kinematics of the wavefronts along the transequatorial loop system is shown in \nfig{fig4} (c), based on which we obtain that the speeds of the incoming and reflected waves were about \speed{897 and -930}, respectively. It should be noted here that only two wavefronts can be identified in the time-distance plot, however, we can observe at least four in the AIA 171 \AA\ images (see \nfig{fig3} and the associated animation). This may be  caused by the weak guiding loop system whose two footpoints are rooted in the high density active regions, while the middle section is off the solar limb where the background density is small. The time-distance stack plot along the eruption direction of the main filament eruption is shown in \nfig{fig4} (d). By fitting one of the helical filament structure with a linear function, it is obtained that the eruption speed was about \speed{215}. 

The oscillations of the transequatorial loop system and the quiescent filament are shown in \nfig{fig4} (e) and (f), respectively. We can track the oscillations for about two cycles, and the trajectories of the loop and filament are marked by white plus signs. By fitting the data points with a damped vibration formula in the form of $D(t) = A e^{-\frac{t}{\tau}} \sin(\frac{2\pi}{T} t+\phi) + B + Ct$, the oscillation parameters can be obtained. Here, $A$, $\tau$, $T$, and $\phi$ in the formula are the amplitude, damping time, period, and initial phase of the oscillation, respectively, and $B$ and $C$ define a linear function that describe the initial position and the linear drifting of the oscillations off the equilibrium position. The fitting results indicate that the oscillation period and amplitude of the loop (filament) oscillation are about 17 (14) minutes and 5.56 (2.58) Mm, respectively. The period of the filament oscillations is well consistent with previous observations and simulations \citep{2012ApJ...745L..18A,2014ApJ...786..151S,2014ApJ...795..130S,2018ApJ...856..179Z}.

The properties of the quasi-periodic wavefront on the solar surface are also investigated by analyzing the variation of the perturbation profiles along the path L1. We show the AIA 193 \AA\  percentage intensity profiles at different times (from 07:43:31 UT to 07:51:55 UT) in \nfig{fig5}(a), from which the first three wavefronts can clearly be identified, and they are indicated by the three red arrows. For each wavefront, the amplitude (intensity increase) first increases to a high value and then gradually decreases to almost the initial undisturbed level. In the meantime, the widths of the wavefronts also showed an increasing trend in time, and the frontal part of each wavefront is much steeper than the rear part. These characteristics suggest the shock nature of the present EUV wave. The largest amplitudes for the first three wavefronts are about 35\%, 23\%, 11\% relative to the background intensity, respectively. Obviously, the amplitude was getting smaller and smaller. Since it is hard to determine the wavelength changes, we just show two intensity profiles at 07:45:07 UT and 07:48:43 UT in panels (b) and (c) of \nfig{fig5}. At this two special moments, both profiles showed two wavefronts simultaneously, and therefore we can determine the wavelengths. Measurements indicate that the wavelength between the first and the second wavefronts was bout 84 Mm, while that between the second and the third ones was about 110 Mm. It seems that the wavefront with a smaller amplitude has a longer wavelength.

To analyze the origin of the quasi-periodic EUV wavefronts, the periodicities of the wavefronts, the helical structures of the unwinding filament, and the associated flare are studied with the wavelet technique \citep{1998BAMS...79...61T}, and the results are plotted in \nfig{fig6}. The detrended intensity profiles along the horizontal white dashed lines as shown in \nfig{fig4} (a) and (b) are used to analyze the periodicities of the EUV wave and the unwinding helical structure of the erupting filament. One can see that the period of the EUV wavefronts along L1 was about $163 \pm 21$ s (\nfig{fig6} (a)), while the period of the unwinding helical structures of the erupting filament was about $150 \pm 13$ s (\nfig{fig6} (b)). The periods in the flare are derived through analyzing the {\em RHESSI} hard X-ray flux in the energy band of 12--25 keV and the flare lightcurves integrated from different AIA wavelength channels. The results indicate that the periods in the flare are in the range of  240--706 s (\nfig{fig6} (c)--(j)). It is interesting that the period of the EUV wavefronts was consistent with the period of the unwinding helical structures of the erupting filament rather than the flare, suggesting that the quasi-periodic EUV wavefronts were possibly excited by the expanding motion of the unwinding helical filament structures.

\section{Seismology Applications}\label{sect4}
Based on the observational results described above, we can estimate the typical magnetic field strength of the closed coronal loop and the large quiescent filament with the method of coronal seismology \citep{2005LRSP....2....3N}. The present peculiar case provides a rare opportunity to estimate the magnetic field strength of the coronal loop via two different ways, and therefore we can test the reliability of these different methods. 

The first method for estimating the magnetic field strength of the coronal loop can use the physical properties of the kink oscillation, using the same method used in \cite{2018ApJ...860...54O}. Based on the AIA 171 \AA\ images, we measured that the width and length of the closed loop system are about $(8.68 \pm 0.36) \times 10^{8}$ cm and $(4.33 \pm 0.64) \times 10^{10}$ cm, respectively. Using the differential emission measure (DEM) inversion code provided by \cite{2015ApJ...807..143C}, the plasma densities of the loop system and the coronal background are estimated to be $(7.96 \pm 0.68) \times 10^8$ $\rm cm^{-3}$ and $(1.57 \pm 0.15) \times 10^{8}$ $\rm cm^{-3}$, respectively. Thus, the plasma density ratio between the background and the loop is $0.20 \pm 0.04$. The kink speed of a loop is defined as $C_{k} = 2L/P$ for the case of the fundamental oscillation mode, here L and P are the loop's length and oscillation period. This reveals that the kink speed is about \speed{849}. On the other hand, the kink speed can also be written as $C_{k} = V_{A}\sqrt{\frac{2}{1 + n_{0}/n_{i}}}$, in which $V_{A} = \sqrt{\frac{B^2}{4\pi\rho}}$ is the Alfv\'en speed inside the transequatorial loop, $n_{0}/n_{i}$ is the plasma density ratio between the coronal background and the loop. Based on these equations, we obtain that the Alfv\'en speed in the loop is about \speed{657}, while the magnetic field strength is about 6.0 Gauss.

The second method for estimating the magnetic field strength of the loop system is by using the physical parameters of the quasi-periodic EUV wave along the loop system. Under the coronal condition where the sound speed is smaller than the Alfv\'en speed, the fast-mode magnetosonic waves propagating along the magnetic field have a velocity of the Alfv\'en speed, i.e., $V_{f} = V_{A}$. The observational results reveal that the average speed of the incoming and reflected waves is $V_{ave}\sim$\speed{914}, If we assume that $V_{ave}$ is just the fast-mode MHD wave speed along the interconnecting loop, the magnetic field strength of the loop system is about 8.3 Gauss. It is noted that the derived magnetic field strength of the coronal loop showed some difference based on the two different methods. We argue that is mainly caused by the shock nature of the present EUV wave, which suggests that we can not simply derive the magnetic field strength using the measured wave speed directly as input. One should firstly divide the measured wave speed by the Mach number, and then derive the magnetic field strength of the guiding field using the real fast-mode wave speed. Alternatively, one can estimate shock Mach number based on the ratio between the observed wave speed $V_{wave}$ and the Alfv\'en speed derived from the kink oscillation of the coronal loop. This should be noted in future studies for estimating magnetic field strength when using the speed of EUV waves.

Considering the EUV wave along the loop system is a shock, its Alfv\'en Mach number $\rm M_A = V_{ave}/V_A$ can be derived to be 1.39. Here, $\rm V_{ave}$ and $\rm V_A$ are the speed of the EUV wave and the Alfv\'en speed in the loop system, respectively. The Mach number of the present EUV wave is consistent with those derived by using other methods in many studies \citep[e.g.,][]{2002ApJ...572L.109N,2003SoPh..212..121H,2012ApJ...752L..23S,2005ApJ...622.1202C,2015LRSP...12....3W}. Generally speaking, the derived magnetic field strength of the loop system based on different methods are all consistent with those obtained by the direct measurement of coronal magnetic field above active regions \citep{2004ApJ...613L.177L,2008ApJ...680.1496L}. 

For simply estimating the radial component of the magnetic field of the remote large quiescent filament, we can use the method proposed by \cite{1966ZA.....63...78H} by using the measured period and damping time of the filament oscillation. This method has been applied in many studies for deriving the filament magnetic field \citep[e.g.,][]{2014ApJ...786..151S,2018ApJ...860..113Z}. In Hyder's method, the relationship between the radial magnetic field and the oscillation parameters is written as $B_{\rm r}^{2} = \pi \rho \, r_{\rm 0}^{2} \, [4 \, \pi ^{2} \, (\frac{1}{P})^{2} + (\frac{1}{\tau})^{2}]$, where $B_{\rm r}$ is the radial magnetic component, $\rho$ is the density of the filament mass, $r_{\rm 0}$ is the scale height of the filament, $P$ is the filament oscillation period, and $\tau$ is the damping time. If we use the value $\rho = 4 \times 10^{-14} \, {\rm g \, cm^{-3}}$, i.e., $n_{\rm e} = 2 \times 10^{10} \, {\rm cm^{-3}}$, the above equation can be rewritten as the form $B_{\rm r}^{2} = 4.8 \times 10^{-12} \, r_{0}^{2} \, [(\frac{1}{P})^{2} + 0.025 \, (\frac{1}{\tau})^{2}]$. Based on this equation and assuming $r_{\rm 0} = 3 \times 10^{9}$ cm \citep{1966ZA.....63...78H}, we can obtain that the radial component of the magnetic field of the filament is about 12.4 Gauss. Based on the inversion of full-Stokes observations, \cite{2003ApJ...598L..67C} published the first map of the vector field in a prominence, and they found that the average magnetic field in prominences is mostly horizontal and varies between 10 to 20 Gauss. This suggests that the estimated magnetic field strength of the filament based on prominence seismology is in the typical range of the results obtained by direct measurement.

\section{Discussions and Conclusions}
By using high spatiotemporal-resolution observations taken by the {\em SDO}/AIA, we report, for the first time, the detailed analysis of an unambiguous quasi-periodic large-scale EUV wave that showed an intensity increase of about 10\%--35\% and can be identified in all EUV channels of {\em SDO}/AIA. The quasi-periodic EUV wavefronts originated from AR11467 successively and were in association with the eruption of a whip-like unwinding eruption filament, a {\em GOES} C3.7 flare, a partial halo CME, and a type II radio burst. The quasi-periodic EUV wavefronts were simultaneously observed along the solar surface and a closed transequatorial loop system that connects AR11467 and AR11469 in the opposite hemispheres. In addition, an upward propagating dome-shaped wavefront off the solar limb was also observed simultaneously during the start of the violent eruption of the  filament. It is interesting that the wavefronts along the transequatorial loop system were reflected back at the southern end of the loop system and caused the obvious kink oscillation of the transequatorial loop system. The quasi-periodic EUV wavefronts along the solar surface in the southeast direction showed refraction effect when they passed through the outskirt of the remote active region AR11469. The interaction between the EUV wave with AR11469 not only launched a jet-like plasma ejecta but also an upward propagating secondary wave above AR11469, which suggests that EUV waves could be the physical linkage of many sympathetic solar eruptions. Furthermore, the quasi-periodic EUV wavefronts in the northeast direction of AR11467 did not propagate to a large distance like those in the southwest direction. This was probably due to the blocking of the magnetic structure hosting the large quiescent filament which showed obvious transverse oscillation after the arrival of the EUV wave.

The interaction part of the EUV wavefront just became more curved and propagated faster than the other part that did not interact with the active region. We interpret this phenomenon as the refraction effect of the EUV wave, which exhibited the real wave nature of the present EUV wave. Earlier simulation and observational works indicated that coronal waves usually tend to be deflected away from regions of high Alfv\'en speeds \citep{1968SoPh....4...30U,2000ApJ...543L..89W,1999ApJ...517L.151T}. However, we do not observe the deflection of the EUV wave away from AR11469 during their interaction. \cite{2013ApJ...773L..33S} reported an interesting EUV wave that simultaneously interacted with two different active regions, in which deflection of the EUV wave just occurred at the boundary of the active region that has a large speed gradient. For the other active region that has a small speed gradient at the boundary, the EUV wave just showed  refraction effect during its passing of the of the active region. Therefore, we propose that no deflection of the present EUV wave during the interaction was probably due to the small speed gradient at the boundary of the active region. Alternatively, if the deflection of the wave did exist, it must be too weak to be detected based on the current AIA observations.

Our measurements indicate that the eruption speed of the active region filament was about \speed{215}. The quasi-periodic EUV wavefronts in the southeast direction were at a nearly  constant speed, whose average speed and deceleration were about \speed{747 $\pm$ 26} and \accel{0.36}, respectively. The period of the quasi-periodic EUV wave was about $163 \pm 21$ seconds. The largest amplitude (intensity increase) of the first three wavefronts were about 35\%, 23\%, and 11\% relative to the background intensity, respectively. This indicates that the amplitudes were getting smaller with each successive wavefronts. In addition, the wavelengths of the quasi-periodic wavefronts were in the range of 84 -- 110 Mm, and it seems that wavefronts with smaller amplitude have a longer wavelength. Although the amplitude of the first three wavefronts showed large differences, their speeds and decelerations are similar. This may indicate that the amplitude do not influence the speed and deceleration of EUV waves significantly. The upward propagating dome-shaped EUV wave component had an initial speed of about \speed{1392} and with a deceleration of about \accel{1.78}. Obviously, the upward propagating dome-shaped EUV wave component was faster and had a stronger deceleration than the wavefronts along the solar surface, which is consistent with previous observational results \citep{2010ApJ...716L..57V}. The speed of the secondary EUV wave above AR11469 was about \speed{1358} and with a deceleration of about \accel{4.5}. Considering the measured parameters of the present EUV waves, they should be the nonlinear large-amplitude waves or shocks as proposed in many statistical studies \citep{2011A&A...532A.151W,2014SoPh..289.4563M,2017SoPh..292..185L}. For the quasi-periodic EUV wavefronts along the transequatorial loop system, the speeds of the incoming and reflected waves were respectively about \speed{897 and -930}, which were faster than that of the wavefronts along the solar surface. The difference between the incoming and reflected waves might be due to the plasma flow inside the transequatorial loop. If this is true, the siphon flow inside the transequatorial loop is from south to the north, with a speed of $\sim$16 km s$^{-1}$. Both the oscillations of the transequatorial loop system and the quiescent filament can be traced for about two cycles, and their amplitude (period) were about 5.56 (2.58) Mm and 17 (14) minutes, respectively. 

A well defined type II radio burst was observed in the RSTO radio dynamic spectra about seven minutes after the first appearance of the EUV wave. Based on the type II radio burst, we derived that the average speed of the shock was about \speed{615} and the start height of the shock was about 0.51 solar radius above the solar surface. It is calculated that the upward propagating dome-shaped wave component off the solar limb had decelerated to a speed of about \speed{644} after seven minutes, consistent with the derived shock speed based on the type II radio burst. This suggests that observed type II radio was caused by the upward dome-shaped wave component. Although quasi-periodic EUV wavefronts were observed along the solar surface and in the transequatorial loop system, whereas there was no quasi-periodic upward propagating dome-shaped wavefronts can be detected in the present case. This could be the reason why there was no quasi-periodic type II radio bursts in the radio dynamic spectra. In addition, this also implies that the periodic excitation of the fast-component EUV wavefronts were mainly visible in the lateral direction of the eruption, or the following wavefronts were too weak to be detected by the AIA observations and therefore can not emit significant radio emissions to be detected by the radio telescope.

Based on the measured parameters of the kink oscillations and the fast-mode EUV wave, two different inversion methods are applied to derive the magnetic field strength of the transequatorial loop system. The results indicate that the Alfv\'en speed in the loop is about \speed{657}, while the magnetic field strength is about 6.0 Gauss. In addition, based on the derived Alfv\'en speed and the measured wave speed, it is obtained that the Mach number of the quasi-periodic shock wave is about 1.39, consistent with the results derived by using other methods in many previous studies. The present peculiar event provides a rare opportunity to evaluate the reliability of different methods of coronal seismology. It is worth noting that the shock nature of the fast-component EUV wave can significantly affect the inversion result when we use the wave speed to derive the magnetic field strength. One should firstly derive the Mach number of the EUV shock wave, and then derive the measured wave speed by the Mach number as the input to derive the magnetic field strength. By using the method proposed by \cite{1966ZA.....63...78H}, the radial component of the filament magnetic field is derived to be about 12.4 Gauss. The values of magnetic field strength of the corona loop and the filament are consistent with the results obtained by direct measurements, suggesting that these inversion methods are reasonable.

The speed of the upward dome-shaped wave component off the solar limb was much faster than those along the solar surface, and the former is about 1.9 times of the latter. \cite{2010ApJ...716L..57V} reported the same characteristic of a limb EUV wave, in which the speed of the upward dome-shaped wave component is about 2.3 times of the surface component. The authors gave two possible explanations for this phenomenon. The first is that the upward dome-shaped wave is possibly driven all the time by the following CME, whereas in the lateral direction the driving force just exists during the initial stage, and then the wave speed is determined by the characteristic speed of the medium; The second is  due to differences of the fast-mode magnetosonic speeds of the active region and the ambient quiet corona; Furthermore, we would like to propose two additional possible explanations for this difference. The first reason is that the fast-mode magnetosonic speed generally increases with height in the low coronal \citep[see,][]{1999ESASP.446..477M,2003SoPh..212..121H}. The second reason could be due to the influence of projection effect, since the outward propagating wave component off the solar limb should suffer from weaker projection effect than the wave component along the solar surface. It is found that the speed of the secondary wave above AR11469 is much higher than the surface component of the primary EUV wave but kinematically compatible with the primary dome component, and both the secondary wave and the dome component showed strong decelerations. Therefore, we propose that the kinematical characteristics of the secondary wave can be explained with the aforementioned possible reasons, in which the height-dependence of the fast-mode magnetosonic speed should be more important than other reasons.

\begin{figure}
\epsscale{0.95}
\figurenum{1}
\plotone{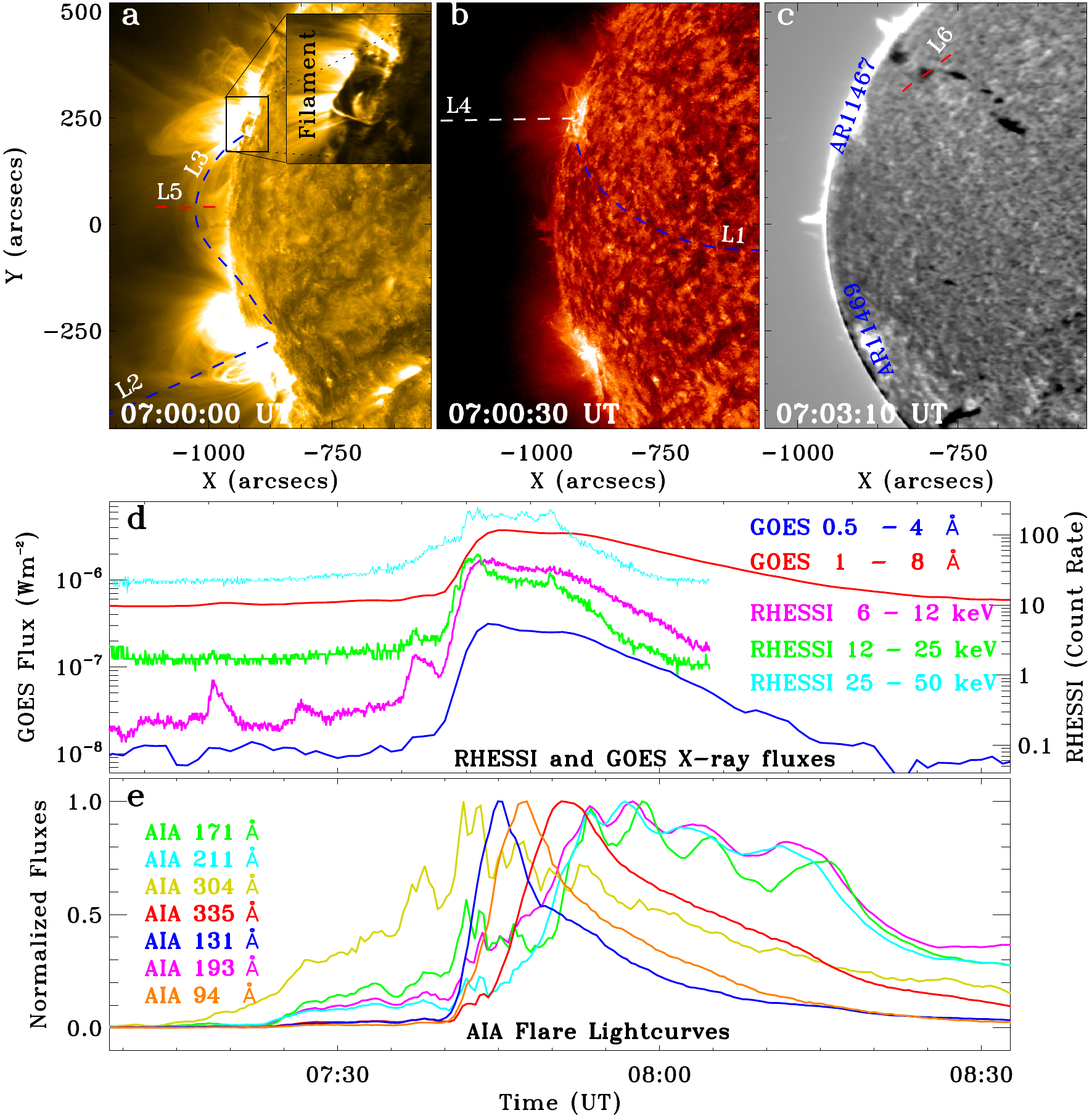}
\caption{An overview of the flare and the initial topology of the eruption source region. Panels (a)--(c) are respectively AIA 171 \AA\, 304 \AA\, and ChroTel H$\alpha$ line center images before the eruption. The inset in panel (a) is the close-up view of the active region filament. Lines L1--L6 are paths used to obtain time-distance stack plots, and the active region numbers are indicated in panel (c). Panel (d) shows the {\em GOES} and {\em RHESSI} X-ray fluxes, in which the different energy bands are plotted in different colors. Panel (e) shows the normalized AIA lightcurves of the eruption source region as shown by the black box in panel (a).
\label{fig1}}
\end{figure}

\begin{figure}
\epsscale{1}
\figurenum{2}
\plotone{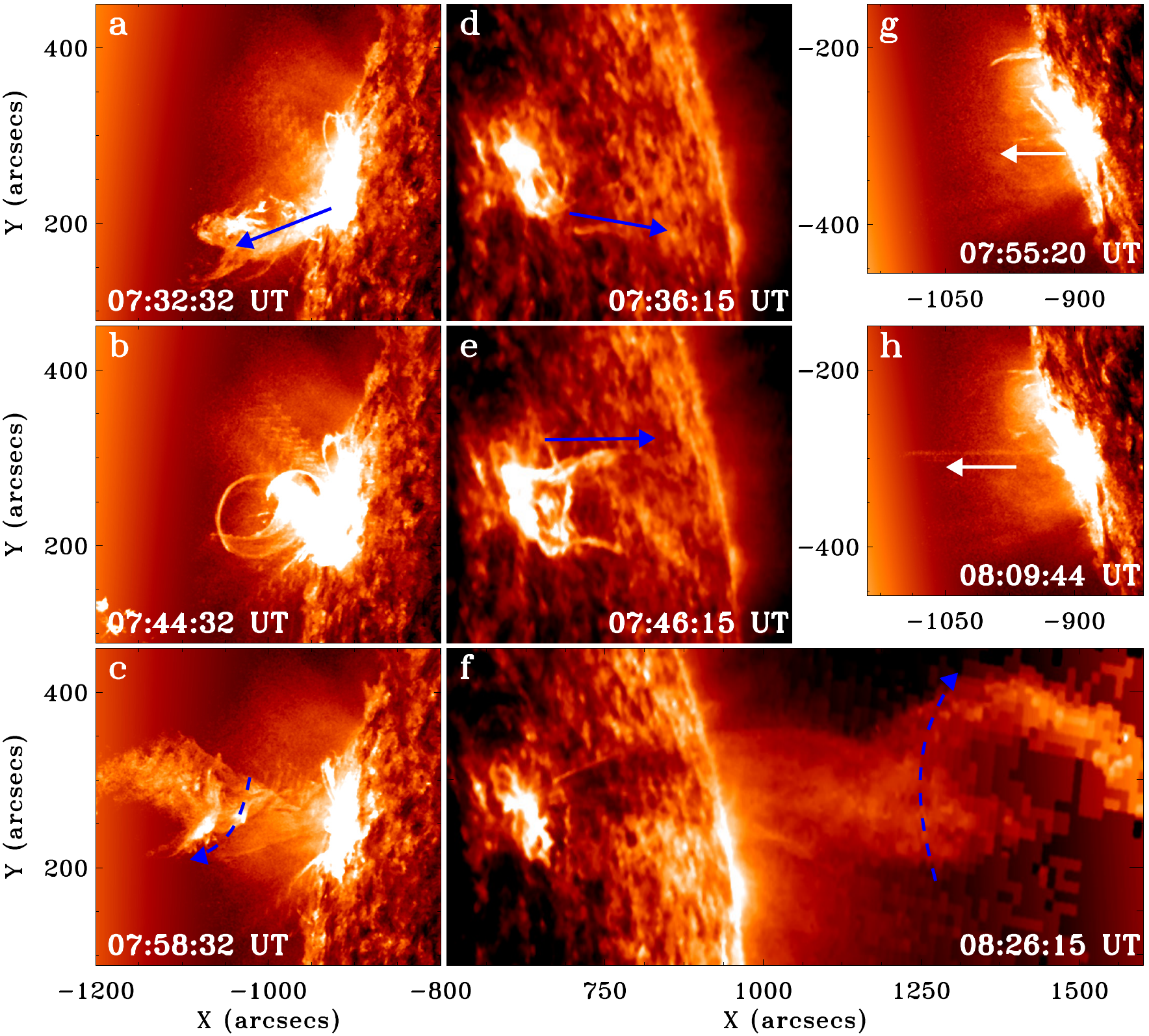}
\caption{The eruption of the whip-like unwinding filament in the eruption source region. Panels (a) -- (c) and (d)--(f) are respectively the AIA and {\em STEREO}-B 304 \AA\ images show the filament eruption in AR11467. The blue arrows in panels (a), (d), and (e) point to the eruption directions of the filament, and the curved arrows in panels (c) and (f) indicate the rotation direction of the unwinding filament. Panels (g) and (h) are AIA 304 \AA\ images show the small plasma ejecta in AR11469, and the white arrows indicate the eruption direction. An animation is available. The top panels show the AIA observations of the AR11467 and AR11469 filament eruptions. These start on 24 April 2012 07:00:08 and end the same day at 08:34:56. The bottom panel is the {\em STEREO}-B observation of AR11467 running from 24 April 2012 70:06:15 to 05:56:15. The duration of the video is 10 seconds.
\label{fig2}}
\end{figure}

\begin{figure}
\epsscale{1}
\figurenum{3}
\plotone{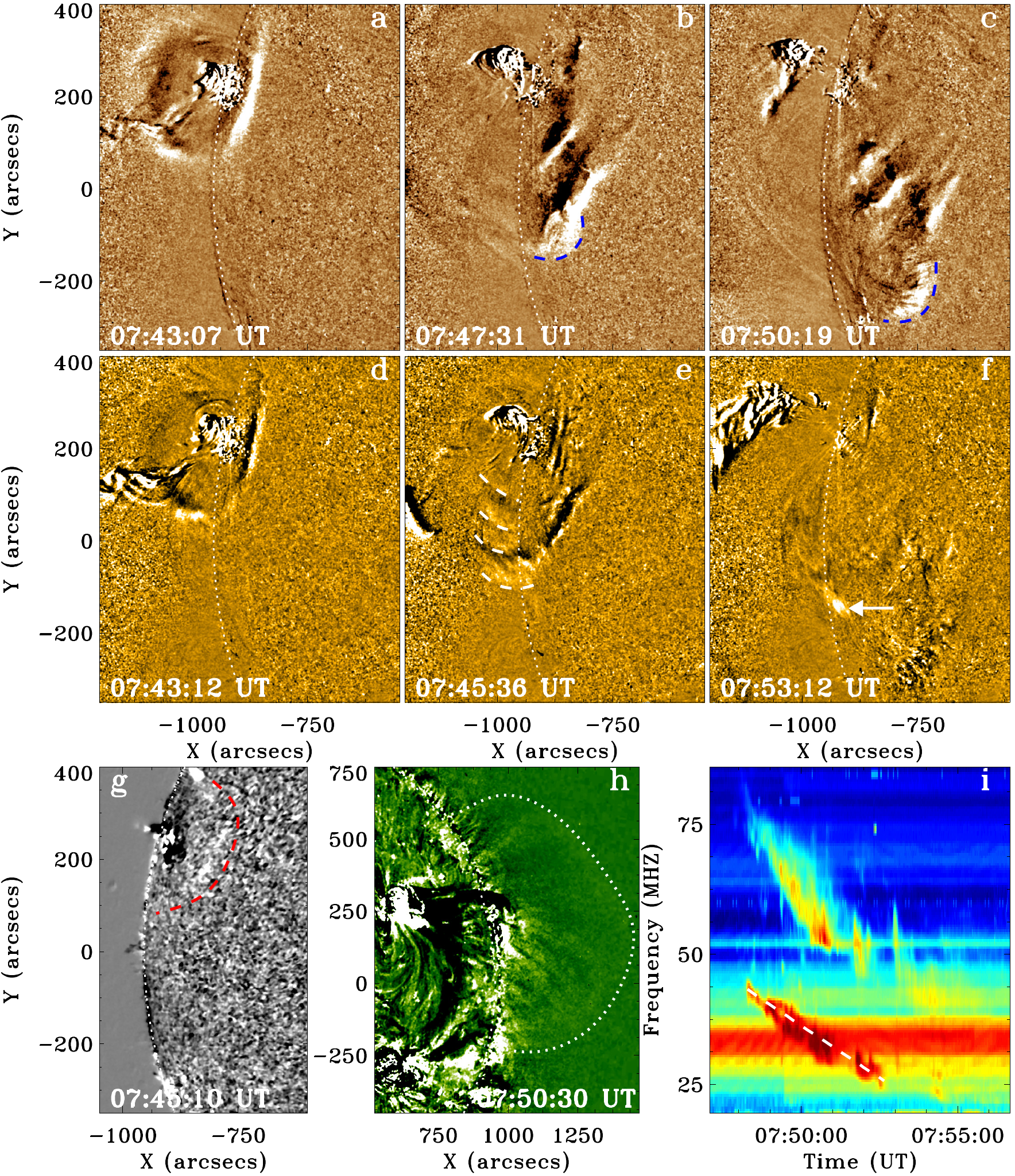}
\caption{The morphology and propagation of the quasi-periodic EUV wave. Top and middle rows are AIA 193 \AA\ and 171 \AA\ running ratio images, respectively. In each panel the white dotted curve indicates the solar limb. The blue dashed curves in panels (b) and (c) indicate the deformation of the first wavefront, while the white dashed curves in panel (e) mark the wavefronts along the closed transequatorial loop system. The white arrow in panel (f) points to the brightening caused by the wave at the loop's southern end. Panel (g) is a ChroTel H$\alpha$ running difference images, and the red dashed curve indicates the Moreton wave. Panel (h) is a {\em STEREO}-B 195 \AA\ running difference images, and the white dotted curve marks the dome-shaped EUV wave. Panel (i) shows the dynamic spectra of RSTO, the white dashed line is a linear fit of the center frequency of the fundamental band of the radio type \uppercase\expandafter{\romannumeral2} burst. An animation showing the AIA 193\AA\ and 171\AA\ direct images and their running ration images. The animation begins on 24 April 2012 07:00:19 and end the same day at approximately 08:20. The video duration is 8 seconds.
\label{fig3}}
\end{figure}

\begin{figure}
\epsscale{0.9}
\figurenum{4}
\plotone{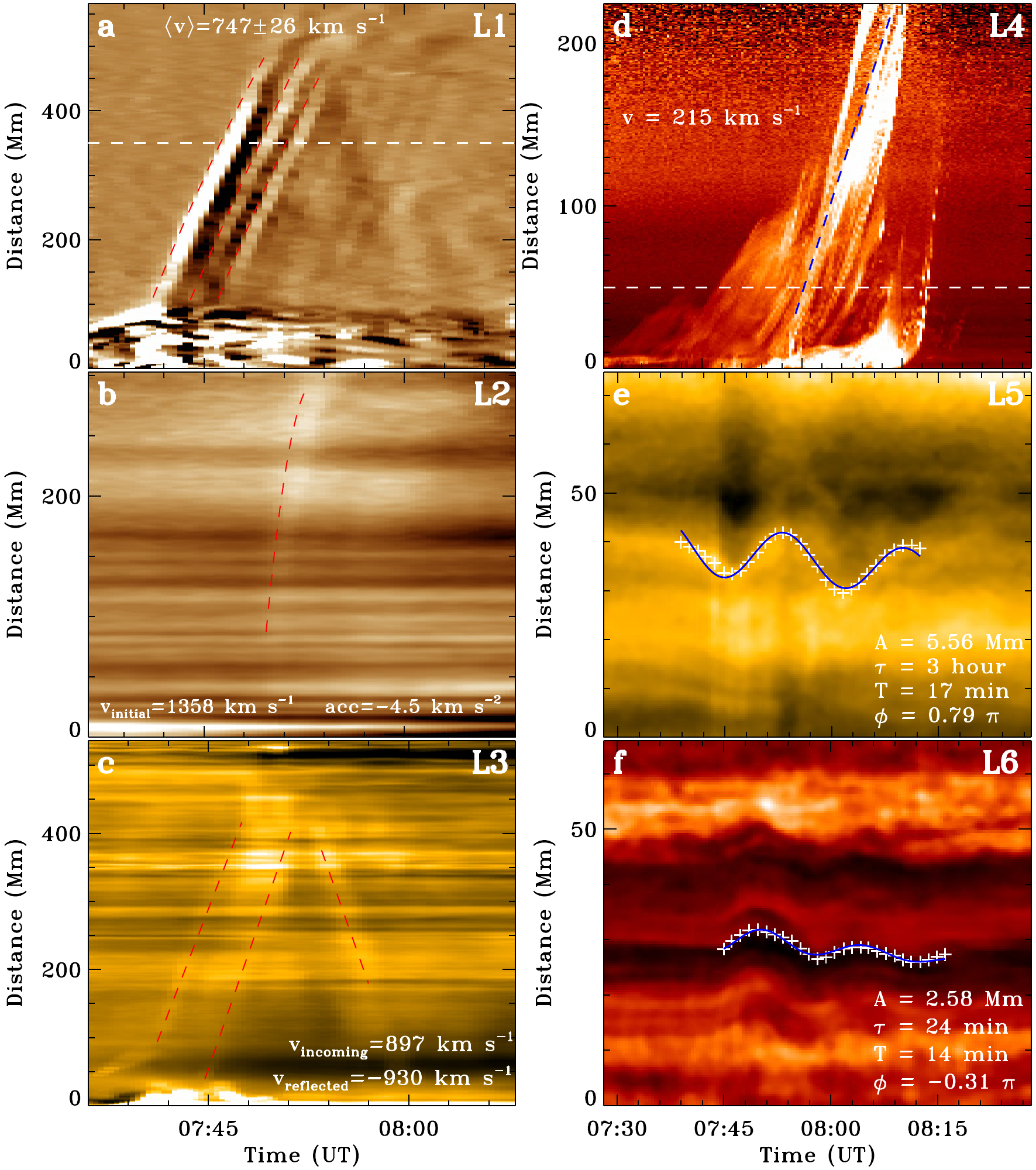}
\caption{Time-distance stack plots along paths L1--L6 as shown in \nfig{fig1}. Panel (a) is made from AIA 193 \AA\ running ratio images, which show the quasi-periodic EUV waves along the solar surface. Panels (b) and (c) are made from AIA 193 \AA\ and 171 \AA\ base ratio images, and they show the waves above AR11469 and along the closed loop system, respectively. Panel (d) shows the eruption of the active region filament. Panels (e) and (f) are made from AIA 171 \AA\ and 304 \AA\ direct images, and they respectively show the transverse oscillations of the closed transequatorial loop system and the quiescent filament. The dashed curves overlaid in panels (a) and (b) are quadratic fit to the wavefronts, while the straight dashed lines in panels (c) and (d) are linear fits to the wavefronts along the closed loop system and the erupting filament. The white horizontal dashed lines in panels (a) and (d) indicate the positions where the intensity profiles are used to analyze the periodicities of the EUV wavefronts and the unwinding helical structures of the erupting filament. The white plus sings in panels (d) and (e) are the measured points of the oscillating loop (filament), while the blue curves are the corresponding fitting results with a damping function. The derived speeds, accelerations, and oscillation parameters are also plotted the figure.
\label{fig4}}
\end{figure}

\begin{figure}
\epsscale{0.95}
\figurenum{5}
\plotone{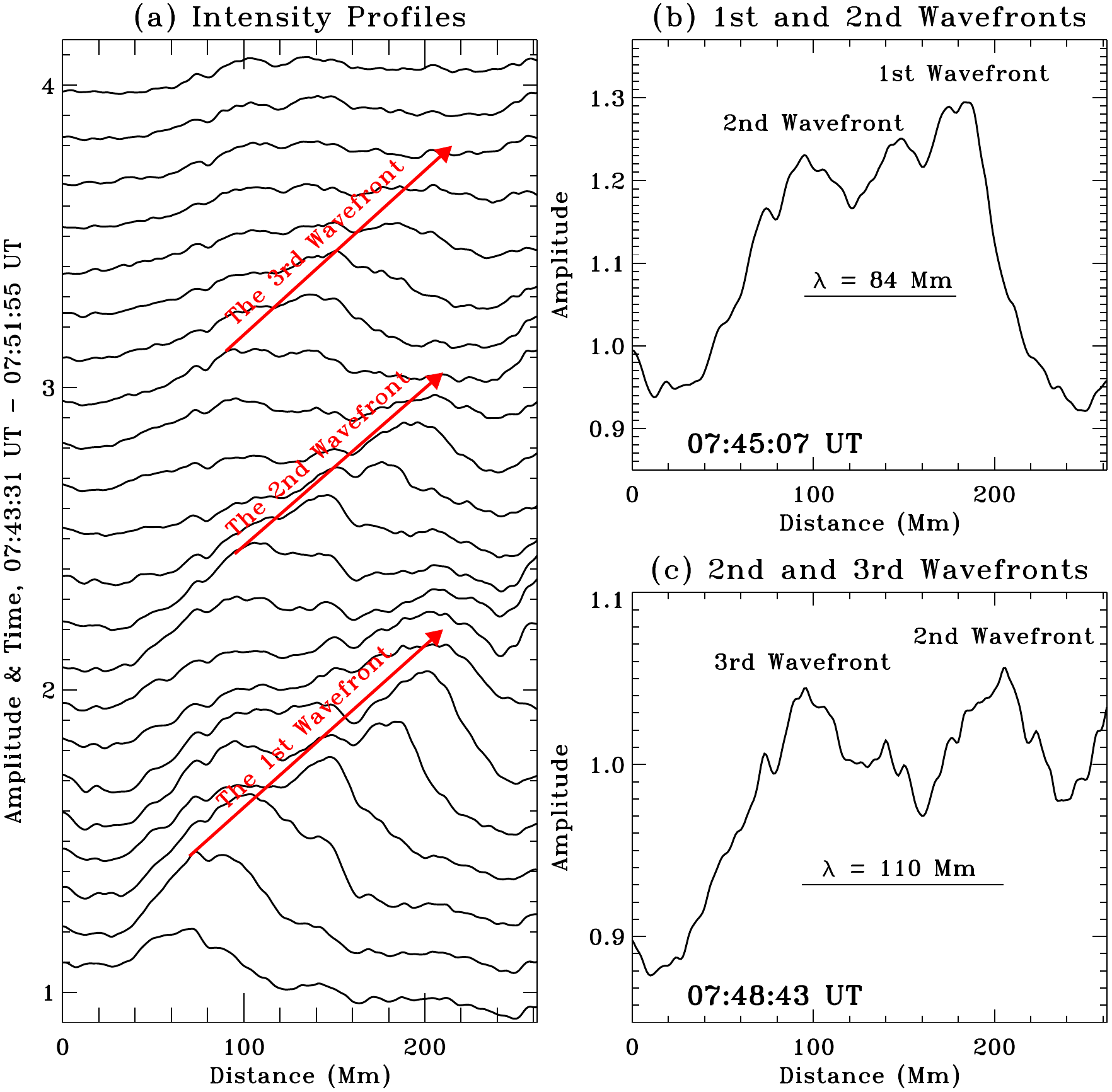}
\caption{Intensity perturbation profiles of the wavefronts along the solar surface in the southeast direction. Panel (a) shows the percentage intensity profiles of the wave at different times, and they are obtained from the AIA 193 \AA\ images. The lowest and the uppermost ones are respectively the intensity profiles at 07:43:31 UT and 07:51:55 UT, and the time interval for neighbouring profiles is 24 second. The red arrows indicate the first three wavefronts. Panels (b) and (c) are the intensity profiles at 07:45:07 UT and 07:48:43 UT, respectively.
\label{fig5}}
\end{figure}

\begin{figure}
\epsscale{0.95}
\figurenum{6}
\plotone{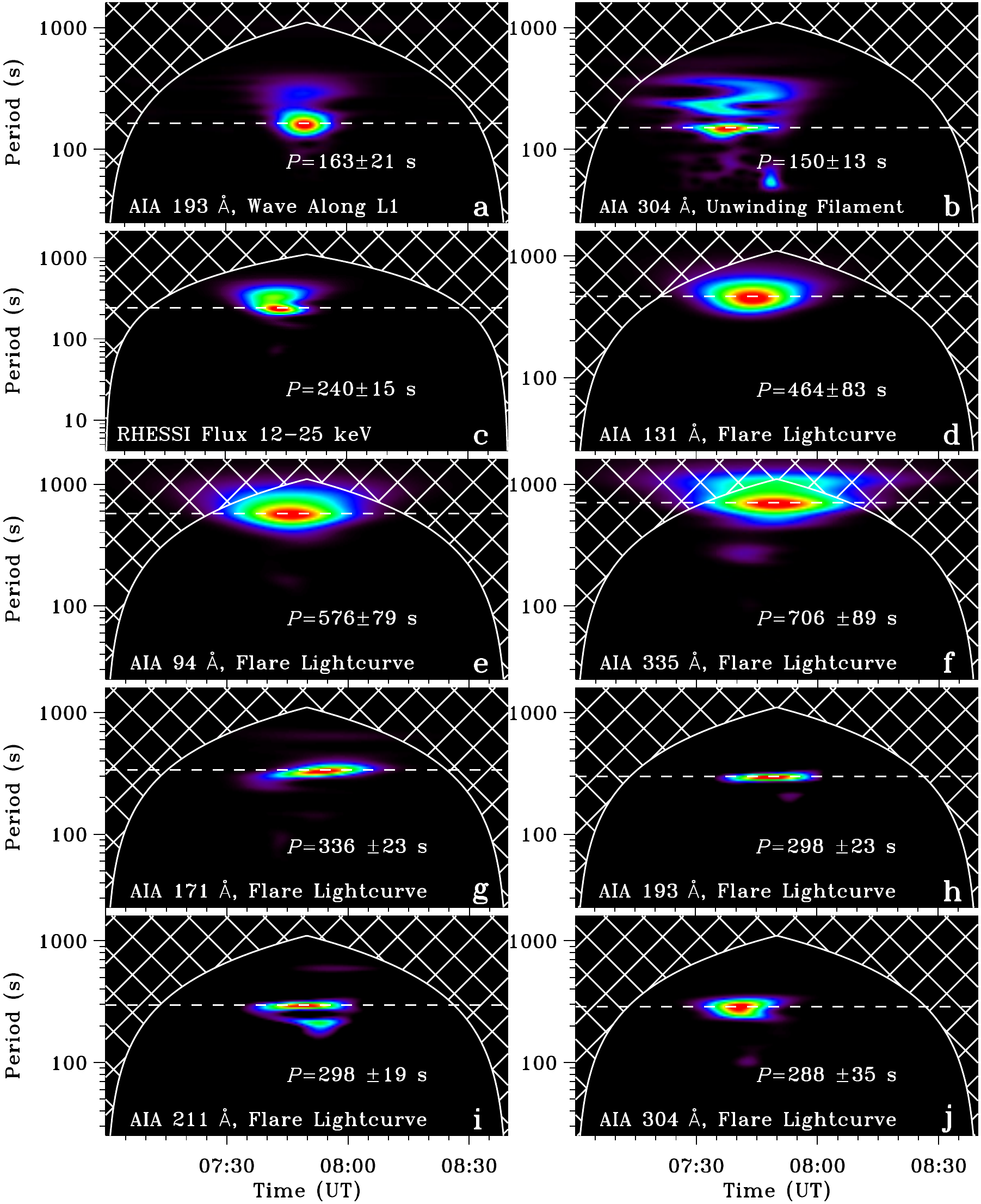}
\caption{Periodicity analysis of the EUV wave, the associated flare, and the unwinding helical filament structure. Panels (a) and (b) are the power maps of the detrended intensity profiles along the horizontal white dashed lines as shown in \nfig{fig4} (a) and (d), respectively. Panel (c) is the wavelet map of the {\em RHESSI} hard X-ray flux in the energy band of 12--25 keV, while panels (d)--(j) are wavelet maps of the AIA flare lightcurves obtained within the flaring region at the AIA's different EUV channels. In each power map, redder color corresponds to higher wavelet power, and the corresponding periods are also indicated in the power maps as horizontal white dashed lines, in which the period errors are given by the half width at half maximum of the power peaks. 
\label{fig6}}
\end{figure}

In most previous observations of EUV waves, only one or two wavefronts can be observed.   For the latter case, the slower wavefront often has a speed of about one third of the preceding faster one, and they can well be interpreted with the hybrid model \citep[e.g.,][]{2002ApJ...572L..99C,2009ApJ...705..587C,2012ApJ...750..134D}. Recent high-resolution AIA observations have demonstrated that sometimes the fast-component EUV wave can consist of two diffuse fronts, with the shallower one leading a broader one \citep{2011SoPh..273..433G}; In addition, the slow-component EUV wave can also consist of multiple small-scale ripple-like fronts, which may represent compression regions ahead of erupting loops \citep{2010ApJ...723L..53L}. The appearance of quasi-periodic wavefronts at a speed of fast magnetosonic wave or shock in EUV should be interesting and important for understanding the formation mechanism and physical properties of EUV waves. The discovery of multiple wavefronts with the fast-mode wave nature can be dated back to \cite{2003SoPh..212..121H}, in which the authors identified the appearance of multiple wavefronts in a few soft X-ray and EUV images. \citet{2004ApJ...610..572G} reported the appearance of five wavefronts in the \ion{He}{1} 10830 \AA\ observations of the 2000 November 25 event. Since their third one was cospatial with the ``EIT wave", according to the hybrid model proposed by \citet{2002ApJ...572L..99C,2005ApJ...622.1202C}, the first two wave fronts should be fast-mode waves. Clear evidence of multiple fast-mode wave fronts came from \citet{2008ApJ...684L..45N}, who showed three Moreton waves in H$\alpha$ images with time intervals being 2.5 minutes and 4 minutes, respectively.  According to Uchida's model, the three Moreton waves are the chromospheric counterparts of fast-mode coronal waves (or shock waves). Their velocities are 580, 1200, and 840 km s$^{-1}$, respectively, which are the typical fast-mode wave speed in the solar corona. They found that the three Moreton waves nicely matched with three partial eruptions of a filament. Therefore, they concluded that filament eruption is indispensable to the generation of Moreton waves, and the three fast-mode waves were generated by successive eruptions of three parts of a filament. Interestingly, large-scale quasi-periodic EUV wavefronts did appear in the numerical simulations presented by \cite{2002ApJ...572L..99C,2005ApJ...622.1202C}. Although not mentioned in the paper, when we look at their \nfig{fig1} of \cite{2002ApJ...572L..99C} and \nfig{fig2} of \cite{2005ApJ...622.1202C}, one can clearly identify the separated multiple wavefronts following the first CME-driven shock wave, with a time interval of 84--168 seconds. Here, the period of the present EUV wave is $163 \pm 21$ seconds, consistent with the upper limit identified in Chen's model. However, it is a pity that the authors did not explain the formation mechanism of these quasi-periodic shocks in their paper. One thing is clear that there is only a single filament eruption in their numerical simulation, hence their quasi-periodicity has nothing to do with repetitive eruptions of filament patches. One possibility is that the multiple fast-mode wave fronts are generated by the oscillation of the magnetic field lines stretched by the erupting flux rope. 

As a similar phenomenon, multiple wavefronts were identified in quasi-periodic propagating fast-mode waves \citep[e.g.,][]{2011ApJ...736L..13L,2012ApJ...753...53S,2013SoPh..288..585S,2018ApJ...853....1S,2014A&A...569A..12N,2017ApJ...851...41Q,2017ApJ...844..149K,2018ApJ...860...54O,2018ApJ...868L..33L,2019ApJ...871L...2M}, which were found to be strongly correlated with the pulses in the solar flare kernels \citep{2005LRSP....2....3N,2015ApJ...807...72L,2017MNRAS.471L...6L,2017A&A...597L...4L}, and the two phenomena may be the different manifestations of a common physical process. However, these waves typically have a narrow temperature range and propagate locally along open or closed coronal loops \citep[see the review paper][and the reference therein]{2014SoPh..289.3233L}.  The quasi-periodic wavefronts observed in the present study are large-scale disturbances with a wide temperature range, and propagated both along and perpendicular to coronal loops. Another difference between our multiple EUV wavefronts and the quasi-periodic propagating fast-mode waves is that the EUV intensity increase of our EUV wave fronts is up to 10\%--35\%, whereas the EUV intensity increase in the quasi-periodic propagating fast-mode waves is only 1\% -- 5\%. According to \cite{Chen16}, the pressure pulse in solar flare kernels tends to generate weak wave fronts with EUV intensity increase of 1\% -- 5\%, whereas the CME-driven (or more precisely, erupting filament-driven) EUV waves are significantly brighter. These observational characteristics spur us to make the conclusion that the observed multiple wavefronts are more like quasi-periodic large-scale fast-mode EUV waves, rather than locally quasi-periodic propagating fast-mode waves \citep[e.g.,][]{2011ApJ...736L..13L,2012ApJ...753...53S,2018ApJ...853....1S}.

To figure out the excitation mechanism of the observed quasi-periodic EUV waves in the present study, the periodicities of the associated flare and the unwinding structure of the erupting filament are analyzed with the generic wavelet software presented by \cite{1998BAMS...79...61T}, and the results indicate that the periods of the associate flare light curve were largely different from the period of the quasi-periodic EUV wave. This suggests that the observed quasi-periodic EUV wave should be excited by other mechanisms rather than the nonlinear physical process in the magnetic reconnection as  quasi-periodic propagating fast-mode waves \citep[e.g.,][]{2011ApJ...736L..13L, 2012ApJ...753...52L,2012ApJ...753...53S,2013SoPh..288..585S,2018ApJ...853....1S}. It is interesting that the period of the quasi-periodic EUV wave is similar to the period of the unwinding helical structure of the whip-like erupting filament in the source active region. Considering the close temporal and period relationships between the quasi-periodic EUV wave and the unwinding motion of the erupting filament, we therefore propose that the observed quasi-periodic EUV waves were excited by the unwinding motion of the erupting helical filament. Here, we do not exclude other possible physical mechanisms, for example, the shaking or oscillation of the erupting filament, the dispersively evolution of an initially broad-band disturbance, and the interaction of the foremost shock with macroscopic inhomogeneous coronal magnetic or plasma structures in or around the source active region. To figure out which physical process excited the observed quasi-periodic wave in the present study, further theoretical and numerical efforts are needed in the future.

\acknowledgments
We thank the excellent observations provided by the {\em SDO}, {\em STEREO}, ChroTel, and RSTO teams. The authors would like to thank the anonymous referee for his/her careful review of the present paper and many valuable comments and suggestions that largely improved the quality of the present work. This work is supported by the Natural Science Foundation of China (11773068,11633008, 11403097, and 11533005), the Yunnan Science Foundation (2015FB191,2017FB006), The West Light Foundation of Chinese Academy of Sciences, and the Specialized Research Fund for State Key Laboratories. P.F.C. was also supported by Jiangsu 333 Project (No. BRA2017359). Y. Shen would like to thank the helpful discussions with Prof. M. Ding in Nanjing University and the ISSI-BJ team members of MHD Seismology of the Solar Corona in the Era of {\em SDO}/AIA.

\end{document}